\documentclass[aps,prl,twocolumn,superscriptaddress,showpacs,floatfix,longbibliography]{revtex4-2}
\pdfoutput=1
\usepackage{amsmath,amssymb,amsfonts,float,graphics,epsfig,epstopdf,color,verbatim,tabularx,bm,multirow,appendix,hyperref}
\usepackage[normalem]{ulem}
\usepackage{lmodern}
\usepackage{ytableau}
\usepackage{pifont}
\usepackage{color}
\usepackage{bm}
\usepackage{wasysym}
\DeclareMathOperator{\Tr}{Tr}
\usepackage{ dsfont }

\def\br{{\mathbf{r}}}

\begin{document}
	
\title{Constraints from anti-unitary symmetries on phase diagrams of sign problem-free models}

\author{Xu Zhang}
\affiliation{Department of Physics and Astronomy, Ghent University, Krijgslaan 281, 9000 Gent, Belgium}

\author{Nick Bultinck}
\affiliation{Department of Physics and Astronomy, Ghent University, Krijgslaan 281, 9000 Gent, Belgium}

\begin{abstract}
A powerful way to guarantee the absence of a sign problem in determinantal quantum Monte Carlo simulations is imposing a particular type of anti-unitary symmetries. It is shown that these same symmetries give rise to constraints on correlation functions, which can be used to identify local operators whose correlations upper bound those of a large class of physically relevant operators, including for example superconducting order parameters. Our results also help understand why it is difficult to realize generic finite-momentum orders in sign problem-free models.
\end{abstract}
\date{\today}
\maketitle

\noindent\emph{Introduction -- } Understanding strongly interacting quantum many-body systems poses a formidable challenge, and the few models that can be solved analytically or numerically are therefore valuable sources of insight. An important question in this context, however, is whether any special physical properties are imposed by requiring a model to be, for example, integrable or sign problem-free. In recent years there has been significant progress in answering this question for the (hydro-)dynamical properties of integrable quantum systems \cite{Vasseur2016,GHD}. Sign problem-free-specific properties seem to be far less understood in general. Some exceptions are given in Refs. \cite{Ringel2017,Ringel2020,Ringel2020_2,Ellison2021}, where the connection between the sign problem and topological phases of matter was discussed, and the recent work~\cite{grossmanrobust2023}, where it was shown that there exists no stable (zero-temperature) Fermi liquid state in a class of sign problem-free Determinantal Quantum Monte Carlo (DQMC) \cite{Scalapino1981,Blankenbecler1981,Hirsch1981} models -- any Fermi liquid in those systems goes superconducting at the lowest temperatures. In this work we uncover a related general constraint on the phase diagrams of sign problem-free DQMC models which crucially rely on an anti-unitary symmetry. In particular, we show that for any of the type of anti-unitary symmetries that can be used to eliminate the sign problem there is a corresponding unique local operator whose Euclidean space-time correlation functions upper bound those of a very general class of local operators. Importantly, many particle-hole and all particle-particle (or pairing) order parameters that have been considered so far in numerical studies of sign problem-free DQMC models belong to this general class. Our result implies that if any of the operators in this class becomes critical, then the local order parameter distinguished by symmetry must become critical as well. Furthermore, the distinguished operators are guaranteed to order at zero momentum, unless the anti-unitary symmetry already encodes a non-zero momentum via a projective commutation relation with translation. Our findings thus put powerful constraints on phase diagrams and critical behavior in sign problem-free DQMC models.

\noindent\emph{Set-up: sign problem-free DQMC models --} We consider a system containing $N$ fermion modes with corresponding creation and annihilation operators $c^\dagger_\alpha,c_\alpha$, $\alpha \in \{1,2,\dots,N\}$. Our results apply to the general class of Hamiltonians of the form $H = H_c[\phi] + H_\phi$. Here, all $c^\dagger_\alpha,c_\alpha$ operators are contained in $H_c[\phi]$, which, moreover, should be quadratic in said operators. For concreteness we will restrict ourselves to models with a fermion number conservation symmetry. The $c$-dependent part of the Hamiltonian also contains additional bosonic operators $\phi$. These bosonic degrees of freedom have their own dynamics which is described by $H_\phi$. Importantly, $H$ has an anti-unitary symmetry $\mathcal{T}$ which is Kramers (i.e. satisfies $\mathcal{T}^2 = -1)$ and acts non-trivially only on the $c^\dagger_\alpha,c_\alpha$ operators. As was shown in Refs ~\cite{wusufficient2005,Lisolving2015,Limajorana2016,lisign2019,weimajorana2016} this provides a sufficient condition for $H$ to be sign problem-free. Note that $\mathcal{T}$ can be a conventional time-reveral symmetry, or a time-reversal symmetry combined with a particle-hole transformation. In the latter case, the anti-unitary symmetry can always be made Kramers by combining it with a suitable action of the U(1) particle number conservation symmetry.  

To derive our main result we will work in the Majorana basis spanned by the operators $\gamma_{1,\alpha} = \frac{1}{\sqrt{2}}(c^\dagger_\alpha+c_\alpha)$ and $\gamma_{2,\alpha} = \frac{i}{\sqrt{2}}(c_\alpha^\dagger-c_\alpha)$. We define $\gamma$ to be a $2N$ ($N\geq1$) dimensional column vector containing the $\gamma_{i,\alpha}$. Without loss of generality, we can take the Kramers anti-unitary symmetries considered in this work to act in the Majorana basis as $\gamma \rightarrow O\gamma$, where $O$ is a $2N\times 2N$ real orthogonal matrix satisfying $O^T= -O$~\cite{supplement}. The anti-unitary symmetry implies that
\begin{equation}
	\Tr(e^{-\beta H} \gamma^{T}(\tau) M \gamma(0)) = \Tr(e^{-(\beta-\tau) H^*} \gamma^{T} e^{-\tau H^*} O^{T} M O \gamma)\,,
\end{equation}
where $M$ is a general $2N\times 2N $ matrix. From this it follows that a general $\gamma$ two-point function $\langle \gamma^{T}(\tau) M \gamma(0) \rangle :=  \Tr(e^{-\beta H} \gamma^{T}(\tau) M \gamma(0))/\Tr(e^{-\beta H})$, satisfies 
\begin{equation}
\langle \gamma^{T} M^* \gamma \rangle=\langle \gamma^{T} O^{T} M O \gamma \rangle^{*}\, . \label{Tprop}
\end{equation}
This property is important for our derivation of the main result in the remainder of the paper.

\noindent\emph{Correlation functions -- } Let us now consider systems with $2^{n+1}$ $\gamma$-modes per spatial unit cell. We define the matrices $M^j$ with $j \in \{1,2,\cdots ,4^{n+1}\}$ as following tensor product basis of the $2^{n+1}\times 2^{n+1}$ matrices: $M^j = \sigma^{\mu_1}\otimes\sigma^{\mu_2}\otimes...\otimes\sigma^{\mu_{n+1}}$, with $\sigma^\mu = (\mathds{1},\boldsymbol{\sigma})$ and $\boldsymbol{\sigma}$ are three Pauli matrices. The local operators we will be considering throughout this work are of the form $\gamma^T(\br) M^j \gamma(\br)$ (note that $M^j$ has to be anti-symmetric), where the spatial index $\br$ labels the unit cells. We will also assume that the matrix $O$ which implements the anti-unitary symmetry on $\gamma$ leaves $\br$ invariant, and is a tensor product of $\mathds{1},\sigma^x,i\sigma^y,\sigma^z$ matrices on the $2^{n+1}$ internal (i.e. intra-unit-cell) degrees of freedom. This implies that $O$ is equal to one of the $M^j$, up to an overall factor that is a power of $i$. For future use, let us also define $\eta_{ij}$ and $\kappa_i$ as follows:
\begin{eqnarray}
M^iM^j & = & (-1)^{\eta_{ij}} M^j M^i \\
OM^i & = & (-1)^{\kappa_i} M^i O
\end{eqnarray}
To start our analysis of the space-time correlators of the bosonic operators $\gamma^T(\tau,\br) M^j \gamma(\tau,\br)$, we first trivially rewrite the thermal average of an operator $P_\gamma$ which is a general polynomial in the $\gamma$ operators but acts trivially on the $\phi$ bosons. Using the imaginary-time path integral formalism we can write
\begin{eqnarray}
\langle P_\gamma \rangle & = & \frac{1}{Z}\int [D\phi][D\gamma] P_\gamma e^{-(S[\gamma,\phi]+S[\phi])} \\
 & = & \frac{1}{Z}\int [D\phi] e^{-S[\phi]}Z[\phi] \left( \frac{1}{Z[\phi]} \int [D\gamma] P_\gamma e^{-S[\gamma,\phi]} \right) \nonumber \\
  & =: & \frac{1}{Z}\int [D\phi] e^{-S[\phi]}Z[\phi] \langle P_\gamma\rangle_{\phi}, \nonumber
\end{eqnarray}
where $S[\gamma,\phi]$ is the part of the imaginary-time action which depends on the $\gamma$ fields and comes from $H_c[\phi]$, and $S[\phi]$ is the part of the action which only depends on the boson fields $\phi$. We have also defined $Z[\phi] = \int[D\gamma]e^{-S[\gamma,\phi]}$, and $\langle P_\gamma\rangle_\phi$ is the expectation value of $P_\gamma$ in a fixed space-time configuration of the boson field $\phi$. Importantly, as the action is quadratic in $\gamma$, we can use Wick's theorem to evaluate $\langle P_\gamma\rangle_\phi$. Furthermore, as the anti-unitary symmetry acts trivially on $\phi$, the action $S[\gamma,\phi]$ is symmetric for every fixed configuration of $\phi$ and this is reflected already in single expectation values $\langle P_\gamma\rangle_\phi$.

Let us now consider the case where $P_\gamma$ is a product of two local $\gamma$ bilinears at two different space-time locations. To streamline our notation, let us from now on use $r=(\tau,\br)$ to label position in both space and imaginary time. Using a Fierz identity~\cite{supplement} and Wick's theorem we can write
\begin{align} \label{FW}
	& \langle \gamma^{T}(r) M^{i} \gamma(r) \gamma^{T}(r') M^{i} \gamma(r')\rangle_\phi   \\
	& = \langle \gamma^{T}(r) M^{i} \gamma(r) \rangle_\phi \langle \gamma^{T}(r') M^{i} \gamma(r')\rangle_\phi \nonumber\\
	& +\frac{1}{2^{n}}
	\sum_{j}(-1)^{\eta_{ij}} \langle \gamma^{T}(r) M^{j} \gamma(r') \rangle_\phi \langle \gamma^{T}(r) M^{j*} \gamma(r') \rangle_\phi. \nonumber
\end{align}
Because the $\gamma$ subsector is $\mathcal{T}$-symmetric for every $\phi$ configuration we can use Eq. \eqref{Tprop} to write
\begin{equation}\label{gam}
\langle \gamma^{T}(r) M^{j*} \gamma(r') \rangle_\phi = (-1)^{\kappa_j} \langle \gamma^{T}(r) M^{j} \gamma(r') \rangle_\phi^*
\end{equation}
Combining Eqs. \eqref{FW} and \eqref{gam} we find
\begin{align} \label{direct}
	& \langle \gamma^{T}(r) M^{i} \gamma(r) \gamma^{T}(r') M^{i} \gamma(r')\rangle_\phi   \\
	& = \langle \gamma^{T}(r) M^{i} \gamma(r) \rangle_\phi \langle \gamma^{T}(r') M^{i} \gamma(r')\rangle_\phi \label{direct} \nonumber\\
	& +\frac{1}{2^{n}}
	\sum_{j}(-1)^{\eta_{ij}+\kappa_j} \big|\langle \gamma^{T}(r) M^{j} \gamma(r') \rangle_\phi\big|^2  \nonumber
\end{align}
Using this result we can write the \emph{connected} correlation function as
\begin{eqnarray}
G^i_c(r-r') & := & \langle \gamma^T(r)M^i\gamma(r) \gamma^T(r')M^i \gamma(r')\rangle \\
 & & - \langle \gamma^T(r)M^i\gamma(r)\rangle\, \langle \gamma^T(r')M^i \gamma(r')\rangle \nonumber\\
  & =: & G^i_{D}(r-r') + G^i_{E}(r-r')\,,
\end{eqnarray}
where the direct part is given by
\begin{align}
 G^i_D(r-r')  =  & \left[\langle \gamma^{T}(r) M^{i} \gamma(r) \rangle_\phi \langle \gamma^{T}(r') M^{i} \gamma(r')\rangle_\phi \right] \\
& - \left[\langle \gamma^{T}(r) M^{i} \gamma(r) \rangle_\phi\right] \left[\langle \gamma^{T}(r') M^{i} \gamma(r') \rangle_\phi \right]\,,\nonumber
\end{align}
with $\left[\,\cdot\, \right] := Z^{-1}\int[D\phi] e^{-S[\phi]}Z[\phi]\left(\,\cdot\,\right) $ the average in the ensemble of space-time configurations of the bosonic field $\phi$. Under very general conditions, this direct part of the connected correlation function can only become critical when $\phi$ itself becomes critical~\cite{supplement}. Moreover, $G^i_D(r-r')$ will be identically zero whenever $\gamma^T(r)M^i\gamma(r)$ has a non-trivial charge under a symmetry of the fermion sector that is preserved for every $\phi$ configuration. An important example of the latter is when $\gamma^T(r)M^i\gamma(r)$ represents a superconducting order parameter. For such operators, critical correlations can only come from the exchange part of the connected correlation function, which is given by
\begin{equation}
G^i_E(r-r') = \frac{1}{2^{n}}
	\sum_{j}(-1)^{\eta_{ij}+\kappa_j} \left[\big|\langle \gamma^{T}(r) M^{j} \gamma(r') \rangle_\phi\big|^2\right]
\end{equation}
It is clear that the exchange part of the connected correlation function will be maximal if we can take $\eta_{ij} = \kappa_j$ for all $j$. This happens iff $M^i \propto O$, which is possible because $
\mathcal{T}$ is Kramers and hence $O$ is anti-symmetric. Also, as the maximal correlation function is strictly positive it is peaked at zero momentum, and hence it can only indicate a critical point where the translation symmetries connecting different unit cells are preserved.

We can now summarize our main result as follows. For every Kramers anti-unitary symmetry acting with $O$ on the Majorana operators $\gamma$ there is a corresponding operator $\gamma^T(r)O\gamma(r)$ which must develop zero-momentum critical correlations at any (zero or finite temperature) second order phase transition in the sign problem-free phase diagram which can be detected with intra-unit cell $\gamma$-bilinear operators, and where the bosonic field $\phi$ is not critical. Note that $\gamma^T(r)O\gamma(r)$ is a superconducting order parameter when $\mathcal{T}$ is a conventional time-reversal symmetry, and a particle-hole bilinear when $\mathcal{T}$ includes a particle-hole transformation \cite{supplement}. In the case of multiple anti-unitary symmetries for which $\langle \gamma^T(r)O\gamma(r)\rangle_\phi$ vanishes by symmetry, the operators $\gamma^T(r)O\gamma(r)$ must always become simultaneously critical, even when this critical point coincides with that of the $\phi$ field. If this happens, then the $\phi$-critical surface is the only critical surface in the phase diagram that can be detected with intra-unit cell $\gamma$-bilinears.

\noindent\emph{Example -- } We now illustrate the power of our result by applying it to the celebrated fermion-boson model introduced in Ref. \cite{Berg2012} to study metallic anti-ferromagnetic critical points~\cite{Berg2019,bauerhierarchy2020}. The model describes spinful fermions hopping on two square lattices labeled by $l = \pm $:
\begin{eqnarray}
	H_c[\phi]&=&\sum_{\br,\br'}\sum_{\sigma=\uparrow,\downarrow}\sum_{l=\pm}c^{\dagger}_{\sigma,l}(\br)t^l_{\br,\br'}c_{\sigma,l}(\br') \\
    &+&\lambda\sum_{\br,l}\sum_{a=x,y,z}(-1)^{r_x+r_y}c^{\dagger}(\br)l^{x}\sigma^{a} c(\br) \phi^{a}(\br)\nonumber
\end{eqnarray}
We can ignore the bosonic part $H_\phi$ as it is not important for our discussion. The coordinates $\br=(r_x,r_y)$ are given by pairs of integers, and for a complete definition of the hopping parameters $t_{\br,\br'}^l$ we refer to Refs.~\cite{Berg2012}. Here we will only require that $t_{\br,\br'}^l$ is non-zero iff it connects the two different sublattices of the square lattice. Defined in this way, the model has two anti-unitary symmetries that act trivially on $\phi$. The first, most obvious, one is a Kramers time reversal symmetry which acts as $l^z\sigma^y$ on the complex fermions. Going to the Majorana basis $\gamma_{i,l,\sigma}$, this time-reversal is implemented by the anti-symmetric matrix $O = l^z i\sigma^y$. We can define a corresponding local hermitian operator $\gamma^T (l^z \sigma^y)\gamma$. Going back to the complex basis, we have $\gamma^T (l^z \sigma^y)\gamma = c^\dagger l^z\sigma^yc^\dagger + h.c.$, which makes it clear that this is a superconducting order parameter. Our general result thus implies that superconductivity must necessarily occur in the $l^z\sigma^y$ channel, as connected correlations in this channel upper bound correlations in all other superconducting channels. The superconducting order must also preserve translation symmetry. This agrees with the numerical results of Refs.~\cite{bauerhierarchy2020}. The second, less obvious, anti-unitary symmetry is a particle-hole symmetry. Combining two neighbouring sites from different sublattices in a single unit cell, we can write the action of the Kramers particle-hole symmetry on the complex fermions as $c^\dagger \rightarrow s^z l^z ic$, where $s^\mu$ act on the intra-unit cell sublattice indices. In the Majorana basis, this action is implemented by the anti-symmetric matrix $O = i\tau^y s^z l^z $, where $\tau^\mu$ act on the $i=1,2$ indices in $\gamma_{i,l,\sigma}$. The corresponding local hermitian operator is given by $\gamma^T \tau^y s^z l^z \gamma = 2c^\dagger s^z l^z c$, which represents a sublattice-staggered layer polarization. Our general result now implies that whenever critical superconducting correlations develop, the staggered layer polarization must also become critical. This follows because the critical correlations must necessarily come from $G_E$ (as $G_D$ is zero for superconducting order parameters), and the two anti-unitary symmetries imply that $G_E$ is identical for both the dominant pairing and staggered layer polarization operators. If we assume that superconductivity indeed develops through a finite-temperature second-order phase transition, then there are two possible scenarios. The first is that the staggered layer polarization develops inside the superconducting region, which would be surprising as the staggered layer polarization and superconducting order parameters are completely unrelated, and from a Landau-Ginzburg perspective there seems to be no reason why they should simultaneously develop long-range order. The second possibility is that the superconducting and layer polarization orders do not coincide, in which case there must be a direct continuous phase transition between them (again under the assumption that superconductivity develops via a second order phase transition). Adding a chemical potential or intra-sublattice hopping breaks the particle-hole symmetry, which is detrimental for the layer polarization, but $l^z\sigma^y$ still remains the dominant pairing channel due to the Kramers time-reversal symmetry.

\noindent \emph{Generalization to models with repulsive interactions -- } Our results can also be generalized to sign problem-free models with repulsive interactions, with one important caveat that is best illustrated via the paradigmatic square (or cubic) lattice  Hubbard model at half filling~\cite{hirschdiscrete1983,hirschtwo1985}:
\begin{equation}
	H=\sum_{\langle \br,\br'\rangle}\sum_{\sigma=\uparrow,\downarrow }c^{\dagger}_{\sigma}(\br)t_{\br,\br'}c_{\sigma}(\br')+U\sum_{\br}(\sum_{\sigma}c^{\dagger}_{\sigma}(\br) c_{\sigma}(\br)-1)^2
\end{equation}
Here $t_{\br,\br'}$ is non-zero only for hopping between the different sublattices. Decoupling the repulsive interaction with a Hubbard-Stratonovich field $\phi$ leads to an effective quadratic Hamiltonian for the fermions. The Hubbard-Stratonovich transformation leaves the hopping term invariant, but introduces following time-dependent and non-hermitian hamiltonian coupling the fermions to $\phi$:
\begin{equation}
	H_I(\tau)=\sum_{\br}i\phi(\tau,\br)(c^{\dagger}_{\uparrow}(\br) c_{\uparrow}(\br)-c_{\downarrow}(\br) c^{\dagger}_{\downarrow}(\br)).
\end{equation}
The quadratic fermion Hamiltonian has an anti-unitary particle-hole symmetry which, because of the spin rotation and charge conservation symmetries, can be defined in several different ways. Again grouping two neighbouring sites belonging to different sublattices in the same unit cell, a first possible definition of the Kramers particle-hole symmetry is $c^\dagger \rightarrow s^z ic$, where $s^z$ again acts on the sublattice index. The corresponding orthogonal transformation in the Majorana basis is $O = i\tau^y s^z$. The hermitian operator associated with this action of the particle-hole symmetry is $\hat{M}_c = \gamma^T \tau^y s^z\gamma = 2c^\dagger s^z c$ (as before, $\tau^y$ acts on the $i=1,2$ index of the Majorana operators). Other possible definitions of the Kramers particle-hole symmetry are $c^\dagger \rightarrow s^z \sigma^{x/z}i c$ and $c^\dagger \rightarrow s^z i\sigma^{y}c$, which respectively act in the Majorana basis as $O^{x/z}=i\tau^ys^z\sigma^{x/z}$ and $O^y = s^z i\sigma^y$. The hermitian operators associated with these symmetries are $\hat{M}_s^{a} = \gamma^T O^a\gamma = 2c^\dagger s^z \sigma^a c$. Physically, $\hat{M}_c$ is the order parameter for $(\pi,\pi)$ charge order, and $\hat{M}_s^a$ are the anti-ferromagnetic order parameters (due to the presence of $s^z$, which gives rise to a staggering on the two different sublattices). Because the fermion Hamiltonian has spin SU(2) symmetry for every $\phi$-configuration, critical spin correlations can only be generated by $G_E$, the exchange part of the correlation function. And as $G_E$ for every intra-unit cell $(c^\dagger,c)$-bilinear is upper bounded by $G_E$ of the $\hat{M}_s^a$ operators, our general result therefore correctly reproduces the well-known fact that the dominant spin-spin correlations are in the anti-ferromagnetic channel. One important subtlety is that our general result also implies that $G_E$ is the same for both $\hat{M}_c$ and $\hat{M}_s^a$, whereas it is physically clear that in the large U limit there will be no critical charge fluctuations at the finite-temperature anti-ferromagnetic transition on the cubic lattice. The resolution here lies in the fact that (1) $\langle \hat{M}_c\rangle_\phi$ is generally non-zero, and (2) the decoupled fermion Hamiltonian is non-hermitian. As a result, $G_D$ will be non-zero for the charge correlations, and is not guaranteed to be positive. In this case, $G_E$, which is strictly positive for the $(\pi,\pi)$ charge order correlations, can be canceled by $G_D$~\cite{supplement}. We emphasize that this can only happen with Hubbard-Stratonovich decouplings that produce non-hermitian fermion hamiltonians.

Let us also briefly mention that our result can be applied in a similar fashion to magic-angle twisted bilayer graphene at charge neutrality. The repulsive Coulomb interaction projected onto the half-filled flat bands also has an anti-unitary particle-hole symmetry which guarantees the absence of a sign problem~\cite{hofmannFermionic2022,zhangMomentum2021,panDynamical2021}. Following the general logic laid out here, this particle-hole symmetry gives rise to a unique local operator with dominant correlations, which exactly corresponds to the Kramers inter-valley coherent order parameter~\cite{bultinckGround2020}.

\emph{Conclusion -- } We have shown how anti-unitary symmetries that guarantee the absence of a sign problem in DQMC simulations lead to constraints on correlation functions which can be used to make general statements about phase diagrams. This was illustrated on several previously-studied models, where we explained how the existing numerical data aligns with our analytical result. In the future we expect our result to be helpful in numerical simulations of new models, with one major application being a priori identification of the dominant superconducting instability in fermion-boson models with Kramers time-reversal symmetry. In general, our results show that before starting DQMC simulations, it is useful to try and identify \emph{all} the anti-unitary symmetries of the model under consideration, as this provides a list of order parameters (some of which might not be obvious from the Hamiltonian or Lagrangian) which have a high chance of developing long-range order somewhere in the phase diagram. Our findings also cast doubt on whether one can find a realistic (i.e. non-contrived) sign problem-free fermion-boson model with a pair density wave instability. Finally, let us also mention that translation symmetry does not enter in an essential way in our arguments, and hence our result is closely related to Anderson's theorem for disordered superconductors \cite{Anderson1959}, and also to a recent extension of this theorem to correlated insulators in magic-angle twisted bilayer graphene (and in particular to the Kramers inter-valley coherent state) \cite{Kolar2023}.

\begin{acknowledgements}
{\noindent \it Acknowledgements.---} This research was supported by the European Research Council under the European Union Horizon 2020 Research and Innovation Programme via Grant Agreement No. 101076597-SIESS (X.Z. and N.B.), and by a grant from the Simons Foundation (SFI-MPS-NFS-00006741-04) (N.B.).
\end{acknowledgements}

\bibliographystyle{apsrev4-2}
\bibliography{Fierz_QMC}

\clearpage
\appendix

\onecolumngrid
\begin{center}
    \textbf{\large Supplementary Material} 
    
    \vspace{0.5 cm}
    
   \textbf{\large Constraints from anti-unitary symmetries on phase diagrams of sign problem-free models}
    
    \vspace{0.5 cm}
    
    Xu Zhang$^{1}$ and Nick Bultinck$^{1}$
    
    \vspace{0.2 cm}
    
    \textit{$^1$Department of Physics and Astronomy, Ghent University, Krijgslaan 281, 9000 Ghent, Belgium}
    
    \vspace{0.5 cm}

    (Dated: \today)
\end{center}

\section{A. Kramers time reversal and anti-unitary particle-hole transformation in Majorana basis}
Here we provide additional details on how to go from the complex fermion to the Majorana fermion basis. Let us define $\Psi$ as the $2N$ dimenstional vector whose first $N$ entries correspond to the annihilation operators $c_\alpha$, and the last $N$ entries correspond to the creation operators $c^\dagger_\alpha$. The unitary transformation which goes from the $\Psi$ basis to the $\gamma$ basis is given by
\begin{equation}
    U=\frac{1}{\sqrt{2}}\begin{pmatrix}
	1 & 1 \\
	  -i & i \\
    \end{pmatrix}\otimes \mathds{1}_N\,,
\end{equation}
i.e. $\gamma = U\Psi$. A Kramers time-reversal which acts on the complex fermions as
\begin{equation}
c^\dagger\rightarrow U_T c^\dagger
\end{equation}
acts on $\Psi$ as
\begin{equation}
\Psi \rightarrow \left(\begin{matrix} U_T^* & \\ & U_T \end{matrix}\right)\Psi
\end{equation}
In the Majorana basis, this becomes
\begin{equation}
\gamma \rightarrow O\gamma =  U^*\left(\begin{matrix} U_T^* & \\ & U_T \end{matrix}\right)U^\dagger \gamma = \frac{1}{2}\left(\begin{matrix} U_T+U_T^* & i(U_T^*-U_T) \\ i(U_T^*-U_T) & -(U_T + U_T^*)  \end{matrix}\right)\gamma
\end{equation}
If $U_T \in \{\sigma^{\mu_1}\otimes\sigma^{\mu_2}\otimes...\otimes\sigma^{\mu_{n}}\}$, then it has to be purely real or purely imaginary, such that $O = U_T \otimes \sigma^z$ or $O = U_T \otimes i\sigma^x$. Moreover, for a Kramers time-reversal symmetry we have that $U_T^T = -U_T$, such that $O$ is indeed anti-symmetric. 

Next let us look at anti-unitary particle-hole symmetries which act as
\begin{equation}
c^\dagger \rightarrow U_P c
\end{equation}
By combining the anti-unitary particle-hole transformation with a U(1) charge symmetry operation we can always ensure that $U_P^2 = -\mathds{1}$, or equivalently $U_P^\dagger = -U_P$. On $\Psi$, the particle-hole transformation acts as
\begin{equation}
\Psi \rightarrow \left(\begin{matrix} U_P^* & \\ & U_P\end{matrix}\right)\Psi^\dagger = \left(\begin{matrix} U_P^* & \\ & U_P\end{matrix}\right) \left(\begin{matrix}  & \mathds{1}\\\mathds{1} & \end{matrix}\right)\Psi
\end{equation}
In the Majorana basis this becomes
\begin{equation}
\gamma \rightarrow O\gamma = U^*\left(\begin{matrix}  & U_P^*\\U_P & \end{matrix}\right) U^\dagger \gamma = \frac{1}{2}\left( \begin{matrix} U_P+U_P^* & i(U_P - U_P^*) \\ i(U_P^*-U_P) & U_P + U_P^*\end{matrix}\right)\gamma
\end{equation}
If $U_P\in \{\sigma^{\mu_1}\otimes\sigma^{\mu_2}\otimes...\otimes\sigma^{\mu_{n}}\}$, then it is either purely real or purely imaginary. In the purely real case we have $O=U_P\otimes \mathds{1}$, and $U_P^T=-U_P$, such that $O$ is indeed real and anti-symmetric. In the purely imaginary case we have $O = U_P \otimes \sigma^y$, and $U_P^T=U_P$ such that again $O$ is real and anti-symmetric.

We summarize these results in Tab. \ref{tab1} below. One can see that in the $\Psi$ basis the two time reversal transformations ($U_T\otimes \mathds{1},U_T\otimes \sigma^z$) give rise to off-diagonal hermitian matrices ($U_T\otimes i\sigma^x,U_T\otimes\sigma^y$) and hence are in the superconducting channel, whereas the two particle-hole transformations ($U_P\otimes\sigma^x,U_P\otimes i\sigma^y$) become diagonal and hence correspond to particle-hole operators ($U_P\otimes i\mathds{1},U_P\otimes i\sigma^z$).

\begin{table}[htp!]
\caption{Translation table for $O=M$ condition between $\Psi$ and $\gamma$ basis}
	\begin{ruledtabular}
		\begin{tabular}{ccccc}
			$U$($\Psi$)&
			$U_T\otimes\mathds{1}$&$U_T\otimes\sigma^z$&$U_P\otimes\sigma^x$&$U_P\otimes i\sigma^y$\\
			\colrule
			$O$($\gamma$)&
			$U_T\otimes\sigma^z$&$U_T\otimes i\sigma^x$&$U_P\otimes\mathds{1}$&$U_P\otimes\sigma^y$\\
			$M$($\gamma$)&
			$U_T\otimes i\sigma^z$&$U_T\otimes\sigma^x$&$U_P\otimes i\mathds{1}$&$U_P\otimes i\sigma^y$\\
			$M$($\Psi$)&
			$U_T\otimes i\sigma^x$&$U_T\otimes\sigma^y$&$U_P\otimes i\mathds{1}$&$U_P\otimes i\sigma^z$\\
		\end{tabular}
	\end{ruledtabular}
	\label{tab1}
\end{table}

\section{B. Fierz identity and correlation function}
Here we give the details of the proof of Eq. \eqref{FW} in the main text.  Recall that the $M^i \in \{\sigma^{\mu_1}\otimes\sigma^{\mu_2}\otimes...\otimes\sigma^{\mu_{n+1}}\}$ are defined as a basis for the $2^{n+1}$-dimensional matrices. For these matrices we can write following Fierz identity:
\begin{equation}
	M_{a,b}^{i}M_{c,d}^{i} = \frac{1}{2^{n+1}}\sum_{j}{(-1)^{\eta_{ij}} M_{a,d}^{j}M_{c,b}^{j}}\,,\label{FierzM}
\end{equation}
where the summation is over all the elements in the set $\{\sigma^{\mu_1}\otimes\sigma^{\mu_2}\otimes...\otimes\sigma^{\mu_{n+1}}\}$, and $M^iM^j = (-1)^{\eta_{ij}}M^jM^i$. This equation can be proved by first noting that we can expand following $2^{n+2}\times 2^{n+2}$ dimensional matrix in the orthogonal basis $\left\{M^j\otimes M^k\right\}$:
\begin{equation}		
	M_{a,b}^{i}M_{c,d}^{i} = {\sum\limits_{j,k}{b_{i;j,k}M_{a,d}^{j}M_{c,b}^{k}}}
\end{equation}
To identify the coefficients $b_{i;j,k}$, we multiply this equation on both sides with $M_{b,c}^mM_{d,a}^n$ and contract the indices to produce
\begin{equation}
	\Tr\left( {M^{i}M^{m}M^{i}M^{n}} \right) = {\sum\limits_{j,k}{b_{i;j,k}\Tr\left( {M^{j}M^{n}} \right)\Tr\left( {M^{k}M^{m}} \right)}}.
\end{equation}
The advantage of the Pauli basis is that the different pairs $M^i$ and $M^j$ always mutually commute or anti-commute, and all the $M^i$ square to the identity. This makes the computation of the trace $\Tr\left( {M^{i}M^{m}M^{i}M^{n}} \right)$ and also $b_{i;j,k}$ straightforward. Using the orthogonality relation $\Tr\left(M^jM^n\right)=2^{n+1}\delta_{j,n}$, we find $b_{i;j,k}=(-1)^{\eta_{ij}}\frac{1}{2^{n+1}}\delta_{j,k}$, from which Eq. \eqref{FierzM} follows. Note that we are only interested in the case where $M^i$ is an anti-symmetric matrix, in which case it follows from Eq. \eqref{FierzM} that
\begin{equation}
M_{a,b}^{i}M_{c,d}^{i} = -\frac{1}{2^{n+1}}\sum_{j}{(-1)^{\eta_{ij}} M_{a,c}^{j}M_{d,b}^{j}}\label{FierzMAS}
\end{equation}
Combining Eqs. \eqref{FierzM}, \eqref{FierzMAS} and Wick's theorem we now find
\begin{align}
	\langle \gamma^{T}(r) M^{i} \gamma(r) \gamma^{T}(r') M^{i} \gamma(r')\rangle_\phi   
	& = \langle \gamma^{T}(r) M^{i} \gamma(r) \rangle_\phi \langle \gamma^{T}(r') M^{i} \gamma(r')\rangle_\phi \\
      &+ M^{i}_{a,b}M^{i}_{c,d}  \bigg[\langle \gamma_{a}(r) \gamma_{d}(r') \rangle_\phi \langle \gamma_{b}(r) \gamma_{c}(r')\rangle_\phi - \langle \gamma_{a}(r) \gamma_{c}(r') \rangle_\phi \langle \gamma_{b}(r) \gamma_{d}(r')\rangle_\phi\bigg] \nonumber\\
    & =  \langle \gamma^{T}(r) M^{i} \gamma(r) \rangle_\phi \langle \gamma^{T}(r') M^{i} \gamma(r')\rangle_\phi \nonumber\\
    + \frac{1}{2^{n+1}}
	\sum_{j}&(-1)^{\eta_{ij}} \bigg[\langle  \gamma^{T}(r) M^{j} \gamma(r') \rangle_\phi \langle \gamma^{T}(r) (M^{j})^{T} \gamma(r') \rangle_\phi + \langle \gamma^{T}(r) M^{j} \gamma(r') \rangle_\phi \langle \gamma^{T}(r) (M^{j})^{T} \gamma(r') \rangle_\phi\bigg] \nonumber\\
    & =  \langle \gamma^{T}(r) M^{i} \gamma(r) \rangle_\phi \langle \gamma^{T}(r') M^{i} \gamma(r')\rangle_\phi  +  \frac{1}{2^{n}}
	\sum_{j}(-1)^{\eta_{ij}} \langle \gamma^{T}(r) M^{j} \gamma(r') \rangle_\phi \langle \gamma^{T}(r) M^{j*} \gamma(r') \rangle_\phi \,,\nonumber
\end{align}
where in the second equality we have used the two different version of the Fierz identity (Eqs. \eqref{FierzM} and \eqref{FierzMAS}), and in the last line we have made use of the hermiticity of the $M^j$. 

\section{C. Criticality of $G_D$}
Let us recall the definition of the direct part of the correlation function:
\begin{equation}
 G^i_D(r-r')  =  \left[\langle \gamma^{T}(r) M^{i} \gamma(r) \rangle_\phi \langle \gamma^{T}(r') M^{i} \gamma(r')\rangle_\phi \right] - \left[\langle \gamma^{T}(r) M^{i} \gamma(r) \rangle_\phi\right] \left[\langle \gamma^{T}(r') M^{i} \gamma(r') \rangle_\phi \right]\,,\nonumber
\end{equation}
where $\langle \gamma^TM\gamma\rangle_\phi$ is the expectation value of the fermionic operator $\gamma^TM\gamma$ in a fixed space-time configuration of the bosonic field $\phi$, and $\left[\,\cdot\, \right] := Z^{-1}\int[D\phi] e^{-S[\phi]}Z[\phi]\left(\,\cdot\,\right) $ represents the average in the ensemble of configurations of $\phi$. We now argue that $G_D$ only becomes critical when $\phi$ becomes critical. Let us first consider the case where the value of the bosonic field is small, such that we can use linear response theory to write
\begin{equation}
\langle \gamma^{T}(r) M^{i} \gamma(r)\rangle_\phi = \int_0^\beta\mathrm{d}\tilde{\tau}\sum_{\tilde{\br}}\chi(\tau-\tilde{\tau},\br-\tilde{\br}) \phi(\tilde{r})\,,
\end{equation}
where $\chi(\tau-\tilde{\tau},\br-\tilde{\br})$ is the imaginary time response function of the unperturbed free fermion hamiltonian, i.e. the free fermion hamiltonian minus the fermion-boson coupling. In this case, $G_D$ becomes
\begin{equation}
G_D^i(r-r') = \sum_{\tilde{r},\tilde{r}'}\chi(r-\tilde{r})\chi(r'-\tilde{r}')\bigg(\left[\phi(\tilde{r}) \phi(\tilde{r}')\right] - \left[\phi(\tilde{r}) \right] \left[\phi(\tilde{r}')\right]\bigg)\,,
\end{equation}
where we have used the short-hand notation $\sum_r = \int\mathrm{d}\tau\sum_{\br}$. If we assume that $\phi$ is not critical, then at long distances we can replace $\left[\phi(\tilde{r}) \phi(\tilde{r}')\right] - \left[\phi(\tilde{r}) \right] \left[\phi(\tilde{r}')\right]$ by an exponentially decaying function $\sim e^{-|\tilde{\tau}-\tilde{\tau}'|/\xi_\tau}e^{-|\tilde{\br}-\tilde{\br}'|/\xi}$. To extract the leading long-distance behaviour we can thus write
\begin{equation}
G_D^i(r-r') \sim \sum_{\tilde{r}}\chi(r-\tilde{r})\chi(r'-\tilde{r})
\end{equation}
The most interesting models have gapless free fermion hamiltonians, in which case the response function can decay algebraically, so let us write $\chi(r-r')\sim |r-r'|^{-\alpha}$. We then find that
\begin{equation}
G_D^i(r-r') \sim \frac{1}{|r-r'|^{2\alpha - d}}\,,
\end{equation}
where $d$ is the space-time dimension. For this to become critical at zero temperature it has to hold that $2\alpha - d < d$, or $\alpha < d$. But this is exactly the condition for the unperturbed free fermion system to be critical at zero temperature, which will only happen under certain fine-tuned conditions that lead to e.g. perfect nesting of a Fermi surface, or a Fermi surface with van Hove singularities. One notable exception are $d=2$ Dirac fermions, which realize the marginal case $\alpha = d$. For higher dimensional Dirac fermions, $\alpha > d$.

For larger values of the bosonic field there will be stronger local pinning of the fermionic expectation values~\footnote{Here we are assuming real electron-boson coupling constants, so \emph{not} coupling to auxiliary fields obtained from Hubbard-Stratonovich decoupling repulsiver interactions.} (especially for values of $\phi(r)$ which are comparable or larger than the fermion hopping strengths), and hence we expect that going beyond linear response theory by writing
\begin{equation}
\langle \gamma^{T}(r) M^{i} \gamma(r)\rangle_\phi = \sum_{\tilde{r}}\chi(r-\tilde{r}) \phi(\tilde{r}) +  \sum_{\tilde{r},\tilde{r}'}\chi^{(2)}(r-\tilde{r},r-\tilde{r'})\phi(\tilde{r})\phi(\tilde{r}') + \cdots
\end{equation}
will not introduce additional critical correlations (i.e. beyond the ones introduced at linear response level for critical free fermion systems), as long as $\phi(r)$ is gapped such that the higher $n$-point functions of $\phi(r)$ also decay exponentially. 

\section{D. Forbidden/Simultaneous criticality}
Here we show that when the bosonic field $\phi$ couples to the fermion bilinear $\gamma^TM^i\gamma$, and the bosonic action is of the form $S[\phi]=\frac{1}{2}\sum_{r}\phi^2(r)$, then:
\begin{itemize}
\item[1)] if the coupling constant is imaginary, both $\gamma^TM^i\gamma$ and $\phi$ cannot become critical,
\item[2)] if the coupling constant is real, $\gamma^TM^i\gamma$ and $\phi$ must become simultaneously critical.
\end{itemize}
Any bosonic field $\phi$ obtained from a Hubbard-Stratonovich decoupling of a fermion interaction term belongs to this class, and this appendix justifies our argument in the main text that for the half-filled Hubbard model $G_D$ of the $(\pi,\pi)$ charge order has to be critical together with $G_E$, such that both cancel each other. 

Let us assume that the electron-boson coupling has the form $i\alpha\phi(r)\gamma^T(r)M^i\gamma(r)$, where $\alpha$ is a real constant. In this case we can write the connected correlation function of $\gamma^TM^i\gamma$ as
\begin{eqnarray}
    &&\langle \gamma^{T}(r) M^{i} \gamma(r)\gamma^{T}(r') M^{i} \gamma(r')\rangle-\langle \gamma^{T}(r) M^{i} \gamma(r)\rangle\langle \gamma^{T}(r') M^{i} \gamma(r')\rangle \\
    &=&\frac{1}{-\alpha^2 Z}\int [D\phi][D\gamma]e^{-S[\phi]} \frac{\partial^2e^{-S[\gamma,\phi]}}{\partial\phi(r)\partial\phi(r')}-\frac{1}{-\alpha^2Z^2}\int [D\phi][D\gamma]e^{-S[\phi]} \frac{\partial e^{-S[\gamma,\phi]}}{\partial\phi(r)}\int [D\phi'][D\gamma']e^{-S[\phi']} \frac{\partial e^{-S[\gamma',\phi']}}{\partial\phi'(r')} \nonumber\\
    &=& -\frac{1}{\alpha^2} \bigg([\phi(r) \phi(r')]-[\phi(r)][\phi(r')]-\delta_{r,r'}\bigg) \label{connG}
\end{eqnarray}
In the last line we have used (functional) integration by parts. The connected correlation functions of both $\gamma^TM^i\gamma$ and $\phi$, interpreted as matrices with row and column indices $r$ and $r'$, must be positive semi-definite (as they are covariance matrices). A critical point corresponds to a diverging eigenvalue of these matrices. Because of Eq. \eqref{connG}, which shows that the eigenvalues $\lambda_\gamma$ of the connected correlation function of $\gamma^TM^i\gamma$ are related to the eigenvalues $\lambda_\phi$ of the connected correlation function of $\phi$ as $\lambda_\gamma = (1-\lambda_\phi)/\alpha^2$, it follows that both $\lambda_\gamma$ and $\lambda_\phi$ must remain finite, implying that both $\gamma^TM^i\gamma$ and $\phi$ cannot become critical.

If the electron-boson coupling has the form $\alpha\phi(r)\gamma^T(r)M^i\gamma(r)$, then the connected correlation function of $\gamma^TM^i\gamma$ can be written as
\begin{eqnarray}
    &&\langle \gamma^{T}(r) M^{i} \gamma(r)\gamma^{T}(r') M^{i} \gamma(r')\rangle-\langle \gamma^{T}(r) M^{i} \gamma(r)\rangle\langle \gamma^{T}(r') M^{i} \gamma(r')\rangle \\
    &=& \frac{1}{\alpha^2} \bigg([\phi(r) \phi(r')]-[\phi(r)][\phi(r')]-\delta_{r,r'}\bigg) \nonumber
\end{eqnarray}
Applying the same logic as above implies that $\phi$ and $\gamma^TM^i\gamma$ must become simultaneously critical.

\section{E. Additional examples}
Finally we give some additional examples of how the physics of well-known sign problem-free DQMC models agrees with our general analytical result.
\subsection{Attractive Hubbard model}
First we consider the flat band attractive Hubbard model~\cite{randeriapairing1992,trivedideviations1995,singerfrom1996,paivacritical2004,tovmasyaneffective2016}. The Hamiltonian is
\begin{equation}
	H=U\sum_{\br}(c^{\dagger}_{\uparrow}(\br) c_{\uparrow}(\br) - c^{\dagger}_{\downarrow}(\br) c_{\downarrow}(\br))^2.
\end{equation}
After a Hubbard-Stratonovich decoupling with an auxiliary field, the fermion Hamiltonian contains the term
\begin{equation}
	H_I(\tau)=\sum_{\br}i\phi(\tau,\br)(c^{\dagger}_{\uparrow}(\br) c_{\uparrow}(\br) - c^{\dagger}_{\downarrow}(\br) c_{\downarrow}(\br)).
\end{equation}
One obvious anti-unitary symmetry of the quadratic fermion Hamiltonian is Kramers time reversal $\sigma^{y}$, which corresponds to the spin-singlet pairing $O_1=c^{\dagger}\sigma^{y}c^{\dagger}+c\sigma^{y}c$. There are two additional anti-unitary particle-hole symmetries which act on the complex fermions as $c^{\dagger}\rightarrow ic$ and $c^\dagger\rightarrow\sigma^{z}ic$, and which give rise to corresponding local operators with maximal $G_E$ that are respectively the charge and spin density. The half-filled attractive Hubbard model has an SU(2) symmetry under which these 3 distinguished operators transform as a vector: $c^\dagger\sigma^yc^\dagger+c\sigma^yc$, $i(c^\dagger\sigma^yc^\dagger-c\sigma^yc)$ and $c^\dagger c$. This SU(2) symmetry implies that these three operators, which all have the same $G_E$, must become simultaneously critical. For the superconducting ones, $G_D$ is zero, which implies that $G_D$ must also be zero for the charge density. At a superconducting critical point, $G_E$ for the spin density also becomes critical. But we expect that $G_E$ will be canceled by $G_D$ according to the discussion of the last section. Introducing a nearest neighbor repulsive/attractive interaction $\mp V(n_i-n_j)^2$ breaks both the SU(2) symmetry, and depending on the sign either the particle-hole or time-reversal symmetry of the decoupled Hamiltonian, which drives system to superconductivity or phase separation~\cite{hofmannsuper2020,hofmannsuper2023}. In a related momentum space projected model, the particle-hole symmetry is broken due to the band dispersion, such that finite temperature superconductivity can exist~\cite{zhang2021superconductivity}.

\subsection{Spin-1/2 Heisenberg model}
As a second example we consider the spin-1/2 Heisenberg model, written in terms of Abrikosov fermions. The nearest-neighbor Heisenberg model on a bipartite lattice is written as
\begin{equation}
	H=J\sum_{\langle \br,\br' \rangle,a=x,y,z}(f^{\dagger}(\br)\sigma^{a}f(\br)) (f^{\dagger}(\br')\sigma^{a}f(\br')) + U\sum_{\br}(f^{\dagger}(\br)f(\br)-1)^2.
\end{equation}
We take $U\rightarrow\infty$ to satisfy the local constraint $f^{\dagger}(\br)f(\br)=1$. Rewriting the $J$ term with a Fierz identity we obtain
\begin{eqnarray}
	H_J&=&-J\sum_{\langle \br,\br' \rangle}[(f^{\dagger}(\br)f(\br')) (f^{\dagger}(\br')f(\br))+h.c.] \nonumber\\
	&=&-\frac{J}{2}\sum_{\langle \br,\br' \rangle}[(f^{\dagger}(\br)f(\br')+f^{\dagger}(\br')f(\br))^2 -(f^{\dagger}(\br)f(\br')-f^{\dagger}(\br')f(\br))^2]
\end{eqnarray}
We have dropped the chemical potential because of the constraint $f^{\dagger}(\br)f(\br)=1$. In the anti-ferromagnetic case with $J>0$, the Kramers anti-unitary particle-hole symmetries of the SU(2)-symmetric free fermion Hamiltonian after Hubbard-Stratonovich decoupling are $f^{\dagger}\rightarrow s^z\sigma^{x/z}i f$ and $f^{\dagger}\rightarrow s^zi\sigma^{y} f$ ($s^z$ again acts on the sublattice index), and in the ferromagnetic case with $J<0$ the Kramers particle-hole symmetries are $f^{\dagger}\rightarrow \sigma^{x/z}i f$ and $f^{\dagger}\rightarrow i\sigma^{y} f$. As expected, this favours criticality of respectively the anti-ferromagnetic order parameter $\hat{M}_{\text{AFM}}^a=f^{\dagger}s^z\sigma^{a}f$, and the ferromagnetic order parameter $\hat{M}_{\text{FM}}^a=f^{\dagger}\sigma^{a}f$.

\subsection{SO(2)$\times$SO(3) Dirac model} 
As a final application we consider the interacting fermion model studied in Refs.~\cite{Ippolitihalf2018,Wangphases2021,Chenphases2024,ZhouSO52024,Chenemergent2024}. We start from the interaction term, which is written in terms of four Dirac spinors $\psi_{\tau,\sigma}$:
\begin{equation}
    H=\frac{1}{2}\int \mathrm{d}^2\br\, U_0[\psi^{\dagger}(\br)\psi(\br)-2]^2-\sum_{i=1}^5u_i[\psi^\dagger(\br)\Gamma^{i}\psi(\br)]^2\,,\label{HIbare}
\end{equation}
with $\Gamma^i\in\{\tau^x\otimes\mathds{1},\tau^y\otimes\mathds{1},\tau^{z}\otimes\sigma^x,\tau^z\otimes\sigma^y,\tau^z\otimes\sigma^z\}$. The parameters are chosen as $u_1=u_2=u_K$ and $u_3=u_4=u_5=u_N$ so as to realize an SO(2)$\times$SO(3) symmetry. To obtain the model studied in Refs.~\cite{Ippolitihalf2018,Wangphases2021,Chenphases2024,ZhouSO52024,Chenemergent2024} one has to project $H$ in the half-filled zeroth Landau level of 4 Dirac fermions. We will perform our analysis directly on Eq.~\eqref{HIbare}, as the Landau-level projection preserves all the anti-unitary symmetries that we will rely on~\cite{Ippolitihalf2018,Wangphases2021,Chenphases2024,ZhouSO52024,Chenemergent2024}. 

Using a Fierz identity one finds that Eq.~\eqref{HIbare} can (up to a constant) be rewritten as
\begin{eqnarray}
    H&=&\int \mathrm{d}^2\br\, g_0[\psi^{\dagger}(\br)\psi(\br)-2]^2+g_1\sum_{a=x,y}[\psi^\dagger(\br)\tau^{a}\psi(\br)]^2   \nonumber\\
    &+&g_2[\psi^\dagger(\br)\tau^z\psi(\br)]^2\,,
\end{eqnarray}
where $g_0=\frac{U+u_N}{2},g_1=-\frac{u_K+u_N}{2}$ and $g_2=u_N$, such that it contains two repulsive and two attractive interactions. Hubbard-Stratonovich decoupling the above Hamiltonian with four auxiliary fields introduces following fermion-boson coupling terms
\begin{align}
H_I&(\tau) = \int\mathrm{d}^2\br\, i\phi^0(\tau,\br)[\psi^\dagger(\br)\psi(\br)-2] \\
& +i\phi^z(\tau,\br)\psi^\dagger(\br)\tau^z\psi(\br) + \sum_{a=x,y}\phi^a(\tau,\br)\psi^\dagger(\br)\tau^a\psi(\br)\,, \nonumber
\end{align}
and hence leads to an SO(3) symmetric quadratic fermion Hamiltonian with anti-unitary particle-hole symmetries acting as $\psi^\dagger\rightarrow\tau^z\sigma^xi\psi$, $\psi^\dagger\rightarrow\tau^z\sigma^zi\psi$, and $\psi^\dagger\rightarrow\tau^zi\sigma^y\psi$, which are all implemented with anti-symmetric matrices in the Majorana basis. Following the general recipe we can construct corresponding local hermitian operators $\hat{M}_{O(3)}^a=\psi^\dagger\tau^z\sigma^a\psi$ which have maximal $G_E$ exchange contributions to the correlation function, and vanishing direct contributions $G_D=0$ due to the SO(3) symmetry. The operators $\hat{M}_{O(3)}^a$ are exactly the SO(3) order parameters.

Our main result now allows us to make following statements about the $(u_N/U,u_K/U)$ phase diagram. At a critical point corresponding to SO(3) symmetry breaking, $G_E$ for the $\hat{M}_{O(3)}^a$ operators becomes critical. Furthermore, at this transition an additional anti-unitary particle-hole symmetry $\psi^\dagger\rightarrow \tau^z i\psi$ implies that $G_E$ is also critical for the local hermitian operator $\psi^\dagger \tau^z\psi$. For this operator, however, $G_D$ is non-zero and will cancel the critical correlations in $G_E$ at the SO(3) critical line~\cite{supplement}. A continuous SO(2)-breaking transition, on the other hand, coincides with a critical $\phi^{x/y}$ field~\cite{supplement}, and we thus find that any putative deconfined quantum critical point between the SO(3) and SO(2) broken regions can be diagnosed by coincident criticality in $G_D$ of both $\tau^z$ and $\tau^{x/y}$ (if this happens, it must occur when $u_N=u_K$ because of the enhanced SO(5) symmetry).

\end{document}